\documentclass[12pt]{article}
\title{Shadow Fields and Local Supersymmetric Gauges}
\usepackage{mathrsfs}
\usepackage{amsmath}

\global\arraycolsep=1pt
\usepackage[colorlinks=true,backref=true,linkcolor=black,anchorcolor=black,citecolor=black,filecolor=black,menucolor=black,pagecolor=black,urlcolor=black]{hyperref}
\usepackage[Symbol]{upgreek}
\usepackage{bbm}
\usepackage{dsfont}
\usepackage{amssymb}
\usepackage{textcomp}
\newcommand{\ba}{/ \hspace{-1.2ex}}
\newcommand{\baa}{/ \hspace{-1.4ex}}
\newcommand{\baaa}{\, / \hspace{-1.6ex}}

\newcommand{\Scal}[1]{\biggl ({#1} \biggr )}
\newcommand{\scal}[1]{\bigl ({#1} \bigr )}
\def\bea{\begin{eqnarray}}
\def\eea{\end{eqnarray}}
\def\be{\begin{equation}}
\def\ee{\end{equation}}
\newcommand{\CR}{\nonumber \\*}

\newcommand{\trace}{\hbox {Tr}~}

\newcommand{\gra}[2]{{\scriptscriptstyle (#1 , #2 )}}
\def\L{{\cal L}}

\DeclareMathAlphabet{\mathpzc}{OT1}{pzc}{m}{it}
\def\s{\,\mathpzc{s}\,}
\def\a{{\scriptscriptstyle (\mathpzc{s})}}
\def\q{{\scriptscriptstyle (Q)}}
\def\qs{{\scriptscriptstyle (Q\mathpzc{s})}}

\def\Lc{\mathscr{L}}

\def\Q{{\mathcal{S}_{\q}}}
\def\phiq{{\varphi^{\q}}}
\def\phiqs{{\varphi^{{\qs}}}}
\def\Omegas{{\Omega^\a}}
\def\Omegaq{\Omega^{\q}}
\def\Omegaqs{\Omega^{{\qs}}}
\def\As{{A^\a}}
\def\Aq{A^{\q}}
\def\Aqs{A^{{\qs}}}
\def\cq{c^{{\q}}}
\def\muq{\mu^{{\q}}}
\def\A{{\scriptscriptstyle A}}
\def\B{{\scriptscriptstyle B}}
\def\C{{\scriptscriptstyle C}}
\def\Ap{{\scriptscriptstyle A'}}

\def\A{a}
\def\B{b}
\def\C{c}
\def\Ap{{a'}}

\def\phis{{\varphi^\a}}

\def\S{{\mathcal{S}_\a}}
\def\P{\mathcal{P}}
\def\F{\mathscr{F}}
\def\G{\overline{{\mathcal{G}}}}
\def\Gs{\overline{{\mathcal{G}_\a}}}
\def\Gq{\overline{{\mathcal{G}_\q}}}
\def\Gqs{\overline{{\mathcal{G}_{\qs}}}}
\def\aG{{\mathcal{G}}}
\def\aGs{{\mathcal{G}^\a}}
\def\aGq{{\mathcal{G}^{\q}}}
\def\aGqs{{\mathcal{G}^{\qs}}}
\def\aGg{{\mathcal{G}^\bullet}}
\def\Gg{\overline{{\mathcal{G}_\bullet}}}
\def\LG{\overline{{L\mathcal{G}}}}
\def\LGs{\overline{{L\mathcal{G}_\a}}}
\def\LGq{\overline{{L\mathcal{G}_\q}}}
\def\LGqs{\overline{{L\mathcal{G}_{\qs}}}}
\def\LaG{L\mathcal{G}}
\def\LaGs{L\mathcal{G}^\a}
\def\LaGq{L\mathcal{G}^{\q}}
\def\LaGqs{L\mathcal{G}^{\qs}}
\def\LaGg{L\mathcal{G}^\bullet}
\def\LGg{\overline{{L\mathcal{G}_\bullet}}}

\def\x{{\mathrm{x}}}
\def\y{{\mathrm{y}}}
\def\R{{\scriptscriptstyle \mathcal{R}}}
\def\zero{{\scriptscriptstyle \rm c}}
\def\un{{\scriptscriptstyle \rm inv}}
\def\im{{\scriptscriptstyle \rm im}}
\usepackage[colorlinks=true,backref=true,linkcolor=black,anchorcolor=black,citecolor=black,filecolor=black,menucolor=black,pagecolor=black,urlcolor=black]{hyperref}

\usepackage[Symbol]{upgreek}
\usepackage{caption}
\usepackage{bbm}
\usepackage{dsfont}
\usepackage{amssymb}
\usepackage{textcomp}
\def\bea{\begin{eqnarray}}
\def\eea{\end{eqnarray}}
\def\be{\begin{equation}}
\def\ee{\end{equation}}

\newcommand\lf[3]{{\mathpzc{l}^{{\scriptscriptstyle (#1,#2)}}_{#3}}}
\newcommand\ff[3]{{\mathpzc{f}^{{\scriptscriptstyle (#1,#2)}}_{#3}}}
\newcommand\gf[3]{{\mathpzc{g}^{{\scriptscriptstyle (#1,#2)}}_{#3}}}
\newcommand\hf[3]{{\mathpzc{h}^{{\scriptscriptstyle (#1,#2)}}_{#3}}}
\newcommand\cf[3]{{\mathpzc{C}^{{\scriptscriptstyle (#1,#2)}}_{#3}}}

\def\L{{\cal L}}

\def\Lc{\mathscr{L}}

\def\susy{{\delta^{\mathpzc{S}}}}
\def\susy{{\delta^{\mathpzc{Susy}}}}

\def\H{{ \mathcal {H} }}
\def\O{{ \mathscr {O} }_{\rm inv}}

\begin{document}
\allowdisplaybreaks[1]
\renewcommand{\thefootnote}{\fnsymbol{footnote}}
\def\corr{$\spadesuit $}
\def\trefle{ $\clubsuit$}
\begin{titlepage}
%\null
\begin{flushright}
CERN-PH-TH/2006-047
% hep-th/
\end{flushright}
\begin{center}
{{\Large \bf
%$\mathcal{N}=4$ SuperYang--Mills\\ from Extended
 % Horizontality Conditions

 Shadow Fields and Local Supersymmetric Gauges
 %\\ and New Local Interactions
 %in Super Yang--Mills Theories (I)% and Renormalized Super Yang--Mills Theories
% Renormalization of super-Yang--Mills in supersymmetric gauge
 }}
\lineskip .75em
\vskip 3em
\normalsize
{\large L. Baulieu\footnote{email address: baulieu@lpthe.jussieu.fr},
 G. Bossard\footnote{email address: bossard@lpthe.jussieu.fr},
 S. P. Sorella\footnote{email address: sorella@uerj.br}}\\
$^{* }$\it Theoretical Division CERN \footnote{ CH-1211 Gen\`eve, 23, Switzerland
}
\\
$^{*\dagger}$ {\it LPTHE, CNRS and Universit\'es Paris VI - Paris VII}\footnote{
4 place Jussieu, F-75252 Paris Cedex 05, France.}
\\
$^{\ddagger}${\it Departamento de F\'\i sica Te\'orica,
%Instituto de F\'\i sica,
UERJ \footnote{Universidade do Estado do Rio de
Janeiro, Rua S\~ao Francisco Xavier 524, 20550-013 Maracan\~a
Rio de Janeiro, Brasil.}}

\vskip 1 em
\end{center}
\vskip 1 em
\begin{abstract}

\end{abstract}

To control supersymmetry and gauge invariance in super-Yang--Mills
theories we introduce new fields, called shadow
fields, which enable us to enlarge the conventional Faddeev--Popov
framework and write down a set of useful Slavnov--Taylor
identities. These identities allow us to address and answer the
issue of the supersymmetric Yang--Mills anomalies, and to perform
the conventional renormalization programme in a fully regularization-independent way.

\end{titlepage}
%%%%%%%%%%%%%%%%%
\renewcommand{\thefootnote}{\arabic{footnote}}
\setcounter{footnote}{0}

%%%%%%%%%%%%%%%%%%%%%%%%%%%%%%%%%%%%%%%%%%%%%%%%%%%%%%%%

%%%%%%%%%%%%%%%%%%%%%%%%%%%%%%%%%%%%%%%%%%%%%%%%%%%%

\renewcommand{\thefootnote}{\arabic{footnote}}
\setcounter{footnote}{0}

%%%%%%%%%%%%%%%%%%%%%%%%%%%%%%%%%%%%%%%%%%%%%%%%%%%%%%%%

%\tableofcontents
 \def\stop{$\blacksquare$}
%%%%%%%%%%%%%%%%%%%%%%%%%%%%%%%%%%%%%%%%%%%%%%%%%%%%%%%%

\section{Introduction}
Renormalizing super-Yang--Mills theories is a subtle issue. One
major difficulty is that of finding the right way of preserving
both supersymmetry and gauge invariance, since supersymmetry
involves the matrix $\gamma_5$ and no consistent regularization
scheme preserving both supersymmetry and gauge invariance is available
so far.

It is worth recalling that, for $\mathcal{N}=1$ super-Yang--Mills
theories, a
superspace formulation is at our disposal.
 In this case, by means of the introduction of superfield
Faddeev--Popov ghosts, the 
 Slavnov--Taylor identity associated to 
super gauge transformations can be established
directly in superspace \cite{piguet}. This has allowed for a superspace
characterization of both counterterms and gauge anomalies, by
means of algebraic methods \cite{piguet,Ferrara:1985me}.
However, the gauge superfield is dimensionless, and its
renormalization in superspace is achieved through non-linear
redefinition and involves an infinite number of parameters in the gauge sector \cite{piguet,nonA}. In the
case of $\mathcal{N}=2,4$ super Yang-Mills theories, a component
field framework, based on the Wess--Zumino gauge, is frequently
employed. Although in this gauge the physical degrees
of freedom are more transparent, gauge and supersymmetric
transformations mix each other in a nonlinear way, a feature that
has required a careful use of the antifield formalism to write
down suitable Slavnov--Taylor identities \cite{slavnov,maggiore}.
However, these identities are not suited for a check of
general supersymmetry covariance: they only control the invariance of observables that are scalar under supersymmetric transformations.

 In this work, these issues will be faced in a novel way. We
introduce new fields in supersymmetric theories,
 which we call shadow fields, not to be confused with the Faddeev--Popov ghosts.
This determines a system of coupled Slavnov-Taylor
identities, which allow for a characterization of the possible
anomalies as well as of the compensating non-invariant
counterterms needed to restore both gauge invariance and
supersymmetry, order by order in perturbation theory,
 in component formalism.
Eventually,
within a class of renormalizable gauges, one gets a proof that
observables truly fall in supersymmetric multiplets.

The method
 improves the Faddeev--Popov gauge-fixing procedure. It
can be applied to all models, for $\mathcal{N}=1,2,4$. Despite their name, these shadow fields add light to the
theory!
Their introduction allows us to write down a family of local gauge-fixing terms with
 several gauge parameters. With some particular choice
of the latter, the gauge-fixing becomes explicitly
supersymmetric, while the result of the usual non-supersymmetric
Faddeev--Popov procedure is recovered for another choice of
the gauge parameters. Both class of gauges determine the same set
of observables, characterized by the cohomology of the BRST
operator. Shadow fields are in fact of great importance, as they
allow us to define the relevant Slavnov--Taylor identities that
control both gauge invariance and supersymmetry, using the
properties of local quantum field theory. 
The possible gauge and supersymmetric anomalies can then be classified. By
introducing the corresponding set of shadow and ghost fields, we thus have
a generalization of the case of the non-supersymmetric Yang--Mills theory,
where Faddeev--Popov ghosts are sufficient.

The content of the
paper is quite general.{\footnote{However, for pedagogical reasons and to
illustrate the method, we sketch at the end the example of
$\mathcal{N}=2$ super-Yang--Mills theory.}} The ordinary BRST
symmetry $\s$ of supersymmetric theories involves the physical
fields and the Faddeev--Popov ghost $\Omega$, the antighost $\bar
\Omega$ and the Lagrange multiplier $b= \s \bar \Omega $. The BRST
doublet $(\bar \Omega, b)$ is cohomologically trivial and is normally used
 to construct a BRST-invariant gauge-fixing term. Observables
are 
 defined as belonging to the cohomology sector
of ghost number zero. Shadow fields are introduced in the form of
two BRST doublets, $(c,\, \mu)$, and $(\bar \mu,\, \bar c)$. They
make it possible to define a second differential operator $Q$ that
consistently anticommutes with $\s$ and carries the relevant
information about supersymmetry. The introduction of the shadow
doublets $(c,\, \mu)$ and $(\bar \mu,\, \bar c)$ is just what is
needed to make the supersymmetric transformations compatible with
a large enough class of local gauge functions. As we shall see,
this allows us to define suitable Slavnov--Taylor identities to 
control both gauge invariance and supersymmetry. This is achieved
by introducing classical external sources for the $\s$, $Q$ and
$\s Q$ transforms of the gauge and matter fields, as well as of
the Faddeev--Popov and shadow fields.
 In fact,
for Green's functions with no external legs of shadow type,
internal loops of shadow fields compensate each other in a way
that is specific to the class of renormalizable gauge that we
choose. Shadow fields can be formally
integrated out by making explicit use of their equations of motion,
 within the Landau gauge. However, this leads to the
appearance of a highly complex and non-local supersymmetry
 of the ordinary Faddeev--Popov action in the Landau gauge,
which cannot be used to obtain meaningful Slavnov--Taylor
identities. Therefore,
 shadow fields have to be kept for non-ambiguously defining the quantization
 of supersymmetric gauge theories.

We also give an explanation of why
 anomalies for supersymmetry and gauge symmetry can
only occur for $\mathcal{N}=1$ models, but not for
$\mathcal{N}=2,\,4$ models. The reason has to do only with the
structure of the
classical supersymmetric algebra, and it boils down to the existence
of non-trivial supersymmetric cocycles. The classification of
anomalies involves only
 the supersymmetric transformations and the set of gauge-invariant
local functional in the physical fields. We generalize
the derivation of the Adler--Bardeen anomaly from a
``Russian formula'', to determine solutions of the consistency conditions
including supersymmetry. This equation allows for an algebraic proof of the
absence of Adler--Bardeen anomaly in extended supersymmetry.
Notice that the demonstration that we display here only holds
for cases where the supersymmetric algebra can be closed without
the use of the equations of motion, i.e. for $\mathcal{N}=1,\,2$. In
the case of $\mathcal{N}=4$ supersymmetry, our method can be used,
but it must be improved. In order to keep a linear dependence on the
sources, one has to restrict to a subalgebra of the whole
supersymmetric one, a question that will be detailed in a
forthcoming paper~\cite{in
 preparation}. Interestingly, shadow
 fields arise very naturally for topological quantum field theories~\cite{in
 preparation}.

%%%%%%%%%%%%%%%%%%%%%%%%%%%%%%%%%%%%%%%%%%%%%%%%%%%%%%%%
\section{Supersymmetry algebra with shadow fields }
%\section{Supersymmetric gauge and Ward identities}
\label{ward} The ``physical'' fields of a generic supersymmetric
 Yang--Mills theory in four dimensions are
the Yang--Mills field $A$ and the matter fields $\varphi^\Ap$,
 which take values in a given representation of the
gauge group. For simplicity, we adopt a notation as if the matter
fields were in the adjoint representation. All our results remain
true when they are in any given representation.\footnote{In the
discussion of the anomalies we will suppose, however, that the
gauge group is semi-simple.} We denote by $\varphi^\A$ the whole
set of fields $(A, \varphi^\Ap)$, and by $\susy$ the generator of
the corresponding supersymmetric transformations. The way two
supersymmetries commute can be generically cast
 in the following form: \be\label{closure} (\susy)^2 \approx
\delta^{\mathrm{gauge}}(\omega(\varphi) + i_\kappa A) + \L_\kappa
\ee where $\approx$ means that this equality can hold modulo
equations of motion. $\susy$ stands for a
supersymmetry transformation, with a ``ghostified''
spinor parameter $\epsilon$, which is commuting. The
quantities $\omega(\varphi) $ and $\kappa$ are bilinear functions
in the parameter $\epsilon$. For example, in the case
of $\mathcal{N}=2$ super-Yang--Mills one has (see section
\ref{exemple})
\be \omega(\varphi) =
\scal{\overline{\epsilon} [ \gamma_5 \phi^5 -
 \phi ] \epsilon } \hspace{10mm} \kappa^\mu = -i
\scal{\overline{\epsilon} \gamma^\mu \epsilon} \ee
For $\mathcal{N}=1,2$, the closure relation (\ref{closure}) can hold
off-shell with the
introduction of auxiliary fields. When auxiliary fields are not
used (for example in $\mathcal{N}=4$ super-Yang--Mills) one has,
for each field $\varphi^\A$, \be (\susy)^2 \varphi^\A + C^{\A\B}
\frac{\delta^L S}{\delta \varphi^\B} =
\delta^{\mathrm{gauge}}(\omega(\varphi) + i_\kappa A) \varphi^\A +
\L_\kappa \varphi^\A \ee where $C^{\A\B}$ are
bilinear functions of the supersymmetric parameters, which do not
depend on the fields.\footnote {This property holds true
for all super-Yang--Mills theories} From the invariance of the
classical action $S$ and from the Jacobi identity for
the differential $\susy$, one has : \begin{gather} C^{\A\B} + (-1)^{\A\B}
C^{\B\A} = 0 \CR \frac{\delta^R \,\susy \varphi^\A}{\delta
\varphi^\C} C^{\C\B} - (-1)^{\A\B} \frac{\delta^R \,\susy
\varphi^\B}{\delta \varphi^\C} C^{\C\A} = 0 \end{gather}

Let $\Omega$ be the Faddeev--Popov ghost for the gauge transformations,
with antighost $\bar\Omega$ and Lagrange multiplier field $b$. The
BRST operator $\s$ is: \be\begin{split}
\s A &= - d_A \Omega\\
\s \Omega &= - \Omega^2\\
\s \bar \Omega &= b
\end{split}\hspace{10mm}\begin{split}
\s \varphi^\Ap &= - [\Omega, \varphi^\Ap]\\
\\
\s b &= 0
\end{split}\ee
%In order to be able to introduce another BRST like operator $Q$ containing
%the supersymmetric transformations,
We introduce two $\s$ trivial doublets, $(\bar \mu,\bar c)$, and
$(c,\mu)$ with:
%in \cite{BGZ} in the introduction of the $w$ symmetry.
\be\begin{split}
\s \bar \mu &= \bar c\\
\s c &= \mu\\
\end{split}\hspace{10mm}\begin{split}
\s \bar c &= 0\\
\s \mu &= 0
\end{split}\ee
We call the set $(\bar \mu^\gra{-1}{-1} ,\, \bar c^\gra{-1}{0},\,
c^\gra{0}{1},\, \mu^\gra{1}{1})$ the shadow quartet. In
 analogy with the Faddeev--Popov ghost number, we
define a shadow number. The upper index $\phi^\gra{g}{s}$ means
that $g$ and $s$ are respectively the ghost and shadow numbers of
the field $\phi$. With this notation, one has for the physical fields
$A ^\gra{0}{0}$ and
 $\varphi^\gra{0}{0}$.
The sum, modulo 2, of $g$, $s$, the
ordinary form degree and the usual Grassmann grading associated to the
 spin, determines the commutation properties of any given field or
 operator. The transformations and the gradings of
the shadow quartet and the Faddeev--Popov ghosts can be deduced from 
the following diagrams: \be\begin{split} &\\* \Omega^\gra{1}{0} \,
& \hspace{8mm} c^\gra{0}{1} \\* &\mu^\gra{1}{1}
\end{split}\hspace{10mm}\begin{split}
&\bar \mu^\gra{-1}{-1}\\*
\bar c^\gra{0}{-1} \, & \hspace{8mm} \bar \Omega^\gra{-1}{0} \\*
&b^\gra{0}{0}
\end{split}\ee
We attribute the values $g=0$ and $s=1$ to each one of the
supersymmetry parameters
 present in the operator $\susy$ (they are commuting spinors). Then, we define a
differential graded operator $Q$ with $g=0$, $s=1$, (for $\s$,
 $g=1$, $s=0$) which represents supersymmetry in a
nilpotent way, as follows: \be\begin{split}
Q A &= \susy A - d_A c \\
Q c &= \omega(\varphi) + i_\kappa A - c^2 \\
Q \Omega &= - \mu - [c, \Omega]\\
Q \bar \mu &= \bar \Omega\\
Q \bar c &= - b
\end{split}\hspace{10mm}\begin{split}
Q \varphi^\Ap &= \susy \varphi^\Ap - [c , \varphi^\Ap]\\
\\
Q \mu &= - [\omega(\varphi), \Omega] - \Lc_\kappa \Omega - [c, \mu]\\
Q \bar \Omega &= \L_\kappa \bar\mu\\
Q b &= - \L_\kappa \bar c
\end{split}\ee
 The operators $\s$ and $Q$ verify
\begin{gather}
\s^2 = 0 \hspace{10mm} Q^2 \approx \L_\kappa \CR
\{\s, Q\} = 0
\end{gather}
These consistent closure relations hold true thanks to the
introduction of the ghost and shadow fields \cite{BZ}. In fact, one
has the graded equation, $(d+\s+Q -i_\kappa)^2=0$, and
 the $\s$ and $Q$ transformations of the fields
can be nicely condensed into a graded horizontality equation
\be\label{hor1} (d + \s + Q - i_\kappa)
\scal{ A + \Omega + c} + \scal{ A + \Omega
 + c}^2 = F + \susy A + \omega(\varphi) \ee
 completed by
 $(d+\s+Q - i_\kappa)\bar \mu =d \mu + \bar c+ \bar\Omega$, and
 their Bianchi identities.\footnote{For any matter super multiplet
 associated to scalar fields $h^\alpha$ in any given representation
 of the gauge group, one has also
\be (d+\s+Q - i_\kappa) h^\alpha + (A + \Omega + c) h^\alpha = d_A
h^\alpha + \susy h^\alpha \nonumber \ee} Eq.~({\ref{hor1}}) is
analogous to the horizontality condition that one encounters in
topological quantum field theories \cite{BBT}. Here, it will be
used for classifying supersymmetric anomalies.
%{\it Je ne suis pas du tout d'accord avec la phrase suivante : }
%These graded horizontality conditions play an important role for
%solving the anomaly equations, as well as for the twisting question
%toward TQFT's

\section {Slavnov--Taylor identities for $\s$ and $Q$ symmetries}

\subsection{Source dependent local effective action}
To extend both $\s$ and $Q$ symmetry at the quantum
level, we must introduce external sources for all $\s$, $Q$ and
$\s Q$ transformations of the fields, which are non-linear. We
 shall consider the following class of gauge-fixing fermion:
\begin{multline}\label{fixing}
\Uppsi = \int \trace \Scal{ \bar\Omega d \star A -
 \frac{\alpha}{2} \star \bar \Omega b + \bar \mu d \star d c + (1-\zeta)
 \bar \mu
 d \star \scal{
 [ A, c] - \susy A} + \frac{\alpha}{2} \star \bar \mu \L_\kappa
\bar c} \\*
= \zeta
 \int \trace \Scal{
 \bar\Omega d\star A -
 \frac{\alpha}{2} \star \bar \Omega b + \bar \mu d \star d c + \frac{\alpha}{2} \star \bar \mu \L_\kappa
\bar c} \\*
+(1-\zeta) \,Q \int \trace \scal{
\bar\mu d\star A- \frac{\alpha}{2}\star \bar\mu b}
 \end{multline}
where $\alpha$ and $\zeta$ are gauge parameters. It leads one to the
following classical gauge-fixed and BRST invariant local
action $\Sigma$, which also contains all needed external sources :
\begin{multline}
\Sigma = S[\varphi] + \s \Uppsi
% \int \trace \biggl( b d\star A - \alpha \star
%b^2 - \bar \Omega d \star d_A \Omega + \bar \mu d \star \scal{ d \mu +
% \zeta [ A, \mu]}\\* + \bar c d \star \scal{ d c + \zeta [ A, c] - \susy
% A } + d\bar \mu \star [ \Omega , \susy A] - \zeta d \bar \mu \star [
%d_A \Omega , c] \biggr)
+ \int (-1)^\A \Scal{ \phis_\A \s \varphi^\A +\phiq_\A Q \varphi^\A +
\phiqs_\A \s Q \varphi^\A}\\
+ \int \trace \Scal{ \Omegas \Omega^2 - \Omegaq Q \Omega - \Omegaqs \s
 Q \Omega + \muq Q \mu - \cq Q c}\\
 + \int \frac{1}{2}\scal{\phiq_\A - [\phiqs_\A, \Omega]} C^{\A\B}
\scal{\phiq_\B - [\phiqs_\B, \Omega]}\\*
+ (1-\zeta) \int \trace ( d \bar c - [\Omega, d\bar \mu]) \star
C^{{ A} \,H^I} \scal{ H^\q_I - [H^\qs_I, \Omega]}\hspace{30mm}
\end{multline}
Here, $H^\q_I$ and $ H^\qs_I$ are the sources for the $Q$ and $\s
Q$ transformations of the possible auxiliary fields $H^I$ of the
chosen supersymmetric model. The last term can be of utility only for
the case $\mathcal {N}=4$.

We will shortly show that the gauge
function $\Uppsi$ is stable under renormalization, for all values
of the parameters $\alpha$ and $\zeta$. This determines a very
interesting class of gauges, which interpolates, in particular,
between the Feynman--Landau gauges, for $\zeta=1$, and a new class
of explicitly supersymmetric gauges, for $\zeta=0$.

%We have employed the
%notation $H^I$ because power counting allows only for auxiliary
%fields of dimension 2.

\subsection{Slavnov--Taylor identity for the $\s$-symmetry}

The $\s$-invariance of the action $\Sigma$ implies the following
Slavnov-Taylor identity, which is valid for all values of
$\alpha$ and $\zeta$: \be \S (\Sigma) = 0 \label{master} \ee
where the Slavnov--Taylor operator $\S$ is given by
\begin{multline}
\S (\F) \equiv \int \Scal{ \frac{\delta^R \F}{\delta \varphi^\A}
 \frac{\delta^L \F}{\delta \phis_\A} + (-1)^\A
 \phiq_\A \frac{\delta^L \F}{\delta
 \phiqs_\A} } \\*
+ \int \trace \Scal{ \frac{\delta^R \F}{\delta \Omega}
 \frac{\delta^L \F}{\delta \Omegas} - \Omegaq\frac{\delta^L
 \F}{\delta \Omegaqs} - \cq \frac{\delta^L
 \F}{\delta \muq} + \mu \frac{\delta^L \F}{\delta c} +
 b\frac{\delta^L \F}{\delta \bar \Omega} + \bar c \frac{\delta^L
 \F}{\delta \bar \mu}}
\label{slavnov}
\end{multline}
 Here and elsewhere, $\F$ is a generical functional of fields and sources. Let us introduce for further use the linearized
Slavnov-Taylor operator
\begin{multline}
\S_{|\F} \equiv \int \Scal{ \frac{\delta^R \F}{\delta \varphi^\A}
 \frac{\delta^L \,\,}{\delta \phis_\A} - \frac{\delta^R \F}{\delta \phis_\A}
 \frac{\delta^L \,\,}{\delta \varphi^\A} + (-1)^\A
 \phiq_\A \frac{\delta^L \,\,}{\delta
 \phiqs_\A} } \\*
+ \int \trace \Scal{ \frac{\delta^R \F}{\delta \Omega}
 \frac{\delta^L \,\,}{\delta \Omegas} - \frac{\delta^R \F}{\delta \Omegas}
 \frac{\delta^L \,\,}{\delta \Omega} - \Omegaq\frac{\delta^L
 \,\,}{\delta \Omegaqs} - \cq \frac{\delta^L
 \,\,}{\delta \muq} + \mu \frac{\delta^L \,}{\delta c} +
 b\frac{\delta^L \,\,}{\delta \bar \Omega} + \bar c \frac{\delta^L
 \,}{\delta \bar \mu}}\label{Lslavnov}
\end{multline}
In particular, the identity (\ref{master}) implies
that ${\S_{|\Sigma}}$ is nilpotent: \be {\S_{|\Sigma}}^2 = 0 \ee
The aim of perturbative gauge theory is to compute, order by order
in the loop expansion, a quantum effective action $\Gamma$ that
satisfies $\S (\Gamma) = 0 $. The property
$\S_{|\F} \S(\F) = 0$ allows us to characterize the possible
gauge anomalies. However, in the present case, one
also has to take into account the $Q$ symmetry, as we will shortly
see.

\subsection{Gauge independence of the $\s$-invariant observables}

The spinor commuting parameters of supersymmetric transformation
$\susy A$ enter the action $\Sigma$ through a $\s$-exact term.
(Notice that these parameters have shadow number~1). As such, they
can be treated as gauge parameters, precisely as the parameters
$\alpha$ and $ \zeta$. This property, combined with the $\s$
invariance of the action, allows us to select observables that do
not depend on the gauge parameters, within the class of gauges
that we are considering.

In fact, the observables are defined
from the cohomology of $\s$, and they turn out to be independent
from these generalized gauge parameters. To prove this property,
one uses the fact that the free propagators and interaction terms
of the theory depend analytically on these gauge parameters, and
thus, the full generating functional also depends analytically
on them. Then, the derivative of the 1PI generating functional
$\Gamma$ with respect to any one of the gauge parameters amounts
to the insertion of a $\s$-exact local term in $\Gamma$. It can be
shown, using for instance the method \cite{PS}, that these insertions
yield a vanishing contribution when inserted in a Green function
of $\s$ invariant observables. Moreover, the theory can be
constructed, while maintaining all relevant Ward identities for
the $Q$ and $\s$ symmetries, for all possible values of the
parameters $\alpha$ and $\zeta$, (Section \ref{nonSusy}). In
fact, the gauges defined by the gauge fermion~(\ref{fixing}) are stable under renormalization, due to some
additional Ward identities stemming from
field equations that are ``linear'' in the quantum fields. As the
Green's functions of $\s$ invariant 
observables are independent on all gauge parameters, their
values are the same within this class of gauges. This gives a nice
way of understanding the supersymmetry covariance of observables,
even for a gauge-fixing that is not manifestly supersymmetric.

When the gauge parameter $\zeta$ is set to 1, the
shadow fields $\bar \mu , \bar c , c$ and $ \mu$ become free,
since, in this case, their contribution in the action is just $\int
\trace (\bar \mu d \star d \mu + \bar c d \star d c)$. The
remaining of the gauge-fixing action is the ordinary
Faddeev--Popov action. In this gauge, supersymmetry is not
manifest, and supersymmetry covariance of observables is very difficult to be established.
 On the other hand, after having introduced the shadows, one 
finds that the Landau--Feynman gauges are a continuous
limit of more general gauges, which are parametrized by $\zeta$
and $\alpha$ and by the parameters of the supersymmetry
transformations. These gauges can now be
made explicitly supersymmetric, by choosing $\zeta = 0$, and
keeping non-zero values for the supersymmetric parameters. Indeed,
in this case, the gauge-fixing term is nothing but a $\s Q$-exact
term, namely \be \label{add} \s Q \int \trace \scal{ \bar \mu d
\star A - \frac{\alpha}{2} \star \bar \mu
 b} \;. \ee
The idea is thus to perform the renormalization programme in the gauge
$\zeta=0$, since in this case the supersymmetric Ward identities
take the simplest form.
%(See section \ref{nonSusy}).
 One relies on the independence theorem of the observables upon changes of the gauge
parameters, so that the result will be the same as in a standard
Faddeev--Popov gauge-fixing, thereby ensuring the supersymmetry
covariance, that is automatic in the gauges where $\zeta=0$,
provided that no anomaly occur for the Ward identities implied by the
$Q$ symmetry.%  Defining the theory for a generic values of $\zeta,\ \alpha$
% is actually not so difficult. As we shall see, there are equations
% of motions that are linear in the quantum fields and allow one to
% directly determine the dependence of the renormalized generating
% functional $\Gamma$ from the fields $b,\bar \Omega, \bar c,\bar
% \mu$, thus characterizing the renormalization properties of the
% gauge fixing-term $\s \Psi$, for all values of $\zeta$ and $ \alpha$. (see Section \ref{nonSusy}).
\subsection{Slavnov--Taylor identity for the $Q$-symmetry in the
 $\zeta=0$ gauge }
In the gauge $\zeta = 0$, as a consequence of the $Q$ and $\s$
invariance of the
gauge-fixing term (\ref{add}), the complete action $\Sigma$
satisfies a second Slavnov--Taylor identity
 \be \Q (\Sigma) = 0
\label{masterQ}\ee where $\Q$ is defined by
\begin{multline}
\Q (\F) \equiv \int \Scal{ \frac{\delta^R \F}{\delta \varphi^\A}
 \frac{\delta^L \F}{\delta \phiq_\A} - (-1)^\A
 \phis_\A \frac{\delta^L \F}{\delta
 \phiqs_\A} - (-1)^\A \L_\kappa \phiqs_\A \frac{\delta^L
 \F}{\delta \phis_\A} + \phiq_\A \L_\kappa \varphi^\A}\\*
+ \int \trace \biggl( \frac{\delta^R \F}{\delta \Omega}
 \frac{\delta^L \F}{\delta \Omegaq} + \frac{\delta^R \F}{\delta c}
 \frac{\delta^L \F}{\delta \cq} +\frac{\delta^R \F}{\delta
 \mu}\frac{\delta^L \F}{\delta \muq} + \Omegas \frac{\delta^L
 \F}{\delta \Omegaqs} + \L_\kappa \Omegaqs \frac{\delta^L
 \F}{\delta \Omegas} \\*+ \Omegaq \L_\kappa \Omega + \cq \L_\kappa
 c + \muq \L_\kappa \mu - b \frac{\delta^L
 \F}{\delta \bar c}- \L_\kappa \bar c \frac{\delta^L
 \F}{\delta b} + \bar \Omega \frac{\delta^L \F}{\delta \bar
 \mu} + \L_\kappa \bar \mu \frac{\delta^L \F}{\delta \bar
 \Omega}\biggr)
\label{slavnovQ}
\end{multline}
Notice that the equation $\Q (\Sigma) = 0$ displays non
homogeneous terms which do not depend on $\Sigma$. However,
the latter are linear in the quantum fields.
One defines a linearized Slavnov--Taylor operator
$\Q_{|\F}$ :
\begin{multline}
\Q_{|\F} \equiv \int \Scal{ \frac{\delta^R \F}{\delta \varphi^\A}
 \frac{\delta^L \,\,}{\delta \phiq_\A}- \frac{\delta^R \F}{\delta \phiq_\A}
 \frac{\delta^L \,\,}{\delta \varphi^\A} - (-1)^\A
 \phis_\A \frac{\delta^L \,\,}{\delta
 \phiqs_\A} - (-1)^\A \L_\kappa \phiqs_\A \frac{\delta^L
 \,\,}{\delta \phis_\A} }\\*
+ \int \trace \biggl( \frac{\delta^R \F}{\delta \Omega}
 \frac{\delta^L \,\,}{\delta \Omegaq}- \frac{\delta^R \F}{\delta \Omegaq}
 \frac{\delta^L \,\,}{\delta \Omega} + \frac{\delta^R \F}{\delta c}
 \frac{\delta^L \,\,}{\delta \cq} -\frac{\delta^R \F}{\delta \cq}
 \frac{\delta^L \,\,}{\delta c} +\frac{\delta^R \F}{\delta
 \mu}\frac{\delta^L \,\,}{\delta \muq}-\frac{\delta^R \F}{\delta
 \muq}\frac{\delta^L \,\,}{\delta \mu}\\* + \Omegas \frac{\delta^L
 \,\,}{\delta \Omegaqs} + \L_\kappa \Omegaqs \frac{\delta^L
 \,\,}{\delta \Omegas} - b \frac{\delta^L
 \,\,}{\delta \bar c}- \L_\kappa \bar c \frac{\delta^L
 \,\,}{\delta b} + \bar \Omega \frac{\delta^L \,\,}{\delta \bar
 \mu} + \L_\kappa \bar \mu \frac{\delta^L \,\,}{\delta \bar
 \Omega}\biggr)
\label{LslavnovQ}
\end{multline}
Provided the identities (\ref{master}) and (\ref{masterQ}) hold
true, the operators $\S_{|\Sigma}$ and $\Q_{|\Sigma}$ satisfy
the following anticommutation relations:
\begin{gather}
{\S_{|\Sigma}}^2 = 0 \hspace{10mm} {\Q_{|\Sigma}}^2 = \P_\kappa \CR
\bigl\{\S_{|\Sigma}, \Q_{|\Sigma}\bigr\} = 0
\end{gather}
 $\P_\kappa$ is the differential operator which acts as the Lie derivative along $\kappa$ on
all fields and
external sources.
\def\z{z}

\vspace{0.5cm}

 \noindent When $\zeta\neq0$, the gauge-fixing term
breaks the $Q$ symmetry. However, this breaking can be kept under
control by introducing an anticommuting parameter $\z$ which forms
a doublet together with the parameter $\zeta$, namely,
$Q\z=\zeta,\ Q\zeta=0$, and $\s\zeta=\s\z=0$. As a consequence,
the term \be
\s \zeta
 \int \trace \Scal{
 \bar\Omega d\star A -
 \frac{\alpha}{2} \star \bar \Omega b + \bar \mu d \star d c + \frac{\alpha}{2} \star \bar \mu \L_\kappa
\bar c}
 \ee
is changed into
 \be \label{Zsource}
\s ( \zeta -\z Q)
 \int \trace \Scal{
 \bar\Omega d\star A -
 \frac{\alpha}{2} \star \bar \Omega b + \bar \mu d \star d c + \frac{\alpha}{2} \star \bar \mu \L_\kappa
\bar c}
 \ee
Thanks to the introduction of the anticommuting parameter $\z$,
the Slavnov--Taylor identity $\Q(\Sigma)=0$ can be maintained for
$\zeta\neq 0$. Section \ref{nonSusy} will be devoted to the study
of this improved Slavnov--Taylor identity, as well as to the
stability of the modified action, with its $\z$ dependence.

 From now on and till section \ref{nonSusy}, we will consider that $\zeta=0$.
 All results, but those involving the ghost Ward identities, can be generalized to the
 case $\zeta\neq 0$, by including the $\z$ and $\zeta$ dependence.
Section~\ref{nonSusy} sketches the relevant modifications.

\subsection{Antighost Ward identities}
\label{antifant}

For all values of $\alpha$, we have ``linear field equations"
for the fields $\bar \mu ,\, \bar c,\, \bar \Omega$ and $b$.
These equations imply the antighost Ward identities that
can be enforced in perturbation theory. They constitute a BRST quartet:
\begin{gather}
\G(\Sigma)=0 \;, \hspace{4mm} \Gs (\Sigma)=0 \;,
\hspace{4mm} \Gq (\Sigma)=0 \;, \hspace{4mm} \Gqs (\Sigma)=0
\end{gather}
where
\begin{gather}
\G(\F) \equiv \int \trace X \Scal{ \frac{\delta^L \F}{\delta b} -
 d\star A + \alpha \star b} \CR
\Gs (\F) \equiv \int \trace X \Scal{\frac{\delta^L
 \F}{\delta \bar \Omega} + d \star \frac{\delta^L \F}{\delta
 \As}} \hspace{10mm} \Gq (\F) \equiv \int \trace X \Scal{\frac{\delta^L
 \F}{\delta \bar c} - d \star \frac{\delta^L \F}{\delta
 \Aq} - \alpha \star \L_\kappa \bar c} \CR
\Gqs (\F) \equiv \int \trace X \Scal{\frac{\delta^L
 \F}{\delta \bar \mu} + d \star \frac{\delta^L \F}{\delta
 \Aqs}}\label{ghost}
\end{gather}
To obtain integrated equations, we introduced a
local arbitrary function $X$, which takes values in the Lie
algebra of the gauge group.
We call the set of the four operators
$\Gg =(\G, \ \Gs,\ \Gq,\ \Gqs)$ the antighost operator quartet.

\subsection{Ghost Ward identities}

For the particular case of
the Landau gauge, $\alpha =0$, we have 
integrated Ward identities:
\begin{gather}
\aGqs (\Sigma)=0 \;, \hspace{4mm}\aGq (\Sigma)=0
\;, \hspace{4mm} \aGs (\Sigma)=0
\end{gather}
with
\bea \aGqs (\F) &=& \int \trace \Scal{ \x \frac{\delta^L
 \F}{\delta \mu} - [\x, \bar \mu] \frac{\delta^L
 \F}{\delta b} + \x \scal{- (-1)^\A [\phiqs_\A, \varphi^\A] +
 [\Omegaqs, \Omega] + [\muq, c]} } \CR
\aGq (\F) &=& \int \trace \biggl( \x \frac{\delta^L
 \F}{\delta c} + [\x, \bar c] \frac{\delta^L \F}{\delta b} -
[\x, \bar \mu] \frac{\delta^L \F}{\delta \bar \Omega} + (-1)^A [\x,
\phiqs_\A] \frac{\delta^L \F}{\delta \phis_\A} - [\x,
\Omegaqs] \frac{\delta^L \F}{\delta \Omegas} \CR
&& \hspace{50mm}
 + \x \scal{[\phiq_\A, \varphi^\A] + [\Omegaq, \Omega] + [\cq, c] +
 [\muq, \mu]}\biggr) \CR
\aGs (\F) &=& \int \trace \biggl( \x \frac{\delta^L
 \F}{\delta \Omega} - [\x, \bar \Omega] \frac{\delta^L
 \F}{\delta b}+ [\x, \bar \mu] \frac{\delta^L \F}{\delta \bar
 c} - [\x, c] \frac{\delta^L \F}{\delta \mu} - (-1)^A [\x, \phiqs_\A]
\frac{\delta^L \F}{\delta \phiq_\A}\CR
&& \hspace{20mm} + [\x, \Omegaqs] \frac{\delta^L \F}{\delta
 \Omegaq} + [\x, \muq]\frac{\delta^L \F}{\delta \cq} + \x [\phis_\A,
\varphi^\A] + \x [\Omegas, \Omega] \biggr) \label{antighost}
\eea $\x$ is an arbitrary constant element of the Lie
algebra. One has also the obvious property \be \aG(\Sigma)=0 \ee
where the linear operator $\aG[\x]$ stands for global gauge
transformations of parameter $-\x$, The operators $\aGg =(\aGqs, \
\aGq,\ \aGs,\ \aG)$ form a quartet. We call $\aGg $ the ghost
operator quartet. % Contrarily to the other Ward identities,
% the ghost Ward identities cannot be generalized to the case $\zeta\neq 0$.

\subsection{Consistency equations for Slavnov--Taylor, antighost and
 ghost operators }

 Given the operators $\Gg$ and $\aGg$, we define their linearized
 functional operators $\LGg$ and $\LaGg$,
 which will complete the linearized
 Slavnov--Taylor operators.\footnote{For example $\LG = \int \trace X
 \frac{\delta^L\,\,}{\delta b}$.} This set of operators builds the
following closed algebra:
\begin{gather}
\S_{|\F} \S(\F) = 0 \hspace{10mm} \Q_{|\F} \Q(\F) = 0 \CR
 \S_{|\F} \Q(\F) + \Q_{|\F} \S(\F) = 0 \CR
\S_{|\F} \G(\F) - \LG \S(\F) = - \Gs(\F) \hspace{10mm} \Q_{|\F} \G(\F)
- \LG \Q(\F) = \Gq(\F) \CR
\S_{|\F} \Gq(\F) + \LGq \S(\F) = \Q_{|\F} \Gs(\F) + \LGs \Q(\F) =
\Gqs(\F) \CR
\S_{|\F} \aGqs(\F) - \LaGqs \S(\F) = - \aGq(\F) \hspace{10mm} \Q_{|\F} \aGqs(\F)
- \LaGqs \Q(\F) = \aGs(\F) \CR
\S_{|\F} \aGs(\F) + \LaGs \S(\F) = \Q_{|\F} \aGq(\F) + \LaGq \Q(\F) =
\aG(\F) \label{consistency1}
\end{gather}
The ghost and antighost operators commute in the following way
\bea \LaGs[\x] \Gs[X](\F) + \LGs[X] \aGs[\x](\F)&=&-\G[[\x,
X]](\F) \CR \LaGq[\x] \Gq[X](\F) + \LGq[X] \aGq[\x] (\F)&=&
\G[[\x, X]](\F) \CR \LaGqs[\x] \Gqs[X](\F) - \LGqs[X] \aGqs[\x]
(\F)&=& \G[[\x, X]](\F) \CR\CR \LaGs[\x] \Gqs[X] (\F) - \LGqs[X]
\aGs[\x](\F)&=&- \Gq[[\x, X]](\F)\CR
 \LaGq[\x] \Gqs[X] (\F) - \LGqs[X] \aGq[\x](\F)&=& \Gs[[\x, X]](\F)\CR
 \\ \nonumber
\LaGs[\x] \aGq[\y] (\F) + \LaGq[\y] \aGs[\x] (\F)&=& -
\aGqs[[\x,\y]] (\F) \CR\\ \nonumber \LaG[\x] \Gg[X](\F) - \LGg[X]
\aG[\x](\F) &=&- \Gg[[\x,X]](\F) \CR \LaG[\x] \aGg[\y](\F) -
\LaGg[\y] \aG[\x](\F) &=&- \aGg[[\x,\y]](\F) \label{consistency2}
\eea where we have written the dependence in the Lie algebra
elements $X$ and x. For simplicity, we shall consider cases
such that \be Q^2 = \L_\kappa \hspace{10mm} C^{\A\B} =
0\ee For instance, by introducing auxiliary fields, this condition
holds true for $\mathcal{N}=1,2$ supersymmetry. We will see how
the case $\mathcal{N}=4$ can be handled in a separate
publication.\cite{ in preparation} 
%%%%%%%%%%%%%%%%%%%%%%%%%%%%%%%%%%%%%%%%%%%%%%%%%%%%%%%%%%%%%%%%%%
\section{Elimination of the shadow fields and non-local
 supersymmetry in the Landau-gauge}
In the class of gauges that we are considering, the
dependence of the action $\Sigma$ on the shadow fields
$c$, $\bar c$ and their ghosts $\bar \mu$ and $\mu$ is quadratic,
including the classical source dependence. Thus, they can be
eliminated by gaussian integration. This defines an
effective action $\Sigma^{\rm eff}$ : \bea\label{simple} && \int
\Biggl . \mathcal{D} c \mathcal{D} \bar c \mathcal{D} \mu
\mathcal{D} \bar \mu \ e^{- \Sigma - \int ( J c
 + \bar J \bar c + K \mu + \bar K \bar \mu )} \Biggr .
 \CR
 &=&
\mathrm{det}^{-1} \left[ \begin{array}{cc} -1 & \alpha G \L_\kappa \\
G \scal{ \mathrm{ad}_{c^\q} + d^\dag \mathrm{ad}_{d_A \Omega} G
 \mathrm{ad}_{\mu^\q} + \mathrm{ad}_{\mu^\q} G
 \mathrm{ad}^\dag_{d_A \Omega} d } & 1 \end{array} \right] \ e^{-\Sigma^{\rm eff}}
\eea The simplicity of the determinant in Eq.(\ref{simple})
 is due to compensations between the
bosonic and fermionic gaussian integrations over the shadow fields
$c,\,\bar c$ and $\mu,\,\bar\mu$. Here, $ \mathrm{ad}_X$ denotes
the adjoint action by a gauge Lie algebra element $X$, and the
symbol $^\dag$ indicates the adjoint of the corresponding
operator with respect to the metric $ \scal{f,g} \equiv \int
\trace \scal{f \star g }$. The operator $G$ is the
inverse of the Faddeev--Popov operator, so that, one has $ G d^\dag d_A =
d^\dag d_A G = 1$.

In the Landau gauge, $\alpha=0$, the
determinant (\ref{simple}) is one. Therefore, in
this gauge, the functional integration over the shadow fields $c$, $\bar
c$, $\bar \mu$ and $\mu$ reduces to the implementation of their
equations of motion. Since the $Q$ transformations depends on $c$ and
$\mu$, the effective $Q$ symmetry then depends on a non-local operator, namely,
on the inverse of the Faddeev--Popov operator. Although perturbation
theory cannot be safely controlled in presence of a symmetry implying
non-local operators, it is interesting to discuss this resulting
effective supersymmetry.

The classical equations of motion of the shadow fields are:
\be\begin{split}
 d_A^\dag d \bar c &\approx J + ... \\*
c &\approx G \scal{ d^\dag \susy A + \bar J}
\end{split}\hspace{10mm}\begin{split}
d_A^\dag d \bar \mu &\approx K + ... \\*
\mu &\approx \s \,G d^\dag \susy A + G \bar K ,
\end{split}\ee
where the dots $...$ stand for terms depending on the
external sources of the $Q$ and $\s Q$ transformations of the
fields. The effective action $\Sigma^{\rm eff}$ is
obtained by substituting the solution of the equations of motion
for $c$, $\bar c$, $\bar \mu$ and $\mu$ in the action $(\Sigma +
\int (J c+ \bar J \bar c + K \mu + \bar K \bar \mu ))$.
 Upon setting the sources $\bar J$ and $\bar K$ to zero,
the resulting effective action $\Sigma^{\rm eff}$ turns out to be
the ordinary Faddeev--Popov action in the Landau gauge, with the
addition of external sources coupled to non-local operators. In
particular, this amounts to replace $c$ by the non
local expression $G d^\dag \susy A$, and $\mu$ by $\s \,G d^\dag
\susy A$. For this effective action, the external sources $J,\,K,\,
c^\q$ and $\mu^\q$ have to be regarded as sources for the non
local operator $G d^\dag \susy A$ and its variations under
$\s,\,Q$ and $\s Q$, respectively.

The corresponding Slavnov--Taylor identities inherit this
non-locality, and become quite complicated. However, one can compute
the 1PI generating functional $\Gamma^{\rm eff}$ associated to the
classical action $\Sigma^{\rm eff}$ in the framework of local quantum
field theory. Let us consider at first the 1PI generating functional $\Gamma$ computed from
the local classical action $\Sigma$ including the shadow fields. Then,
we eliminate the external shadow fields by Legendre transformation, substituting to them their
solutions $c^*,\,\mu^*,\, \bar c^*$ and $\mu^*$ of the
equations:\footnote{These equations are solvable as power series in $\hbar$ since the
  classical equations of motions are solvable.} 
\be\begin{split} \frac{\delta^R
\Gamma}{\delta c} &= - J
\\* \frac{\delta^R \Gamma}{\delta \bar c} &= 0
\end{split}\hspace{10mm}\begin{split}
\frac{\delta^R \Gamma}{\delta \mu} &= - K \\*
\frac{\delta^R \Gamma}{\delta \bar \mu} &= 0
\end{split}\label{qE}\ee
These solutions are non-local functionals of the sources $J$ and $K$, and of other
fields. The Legendre transformation,
\be
\Gamma^{\rm eff}[J,K] \equiv \Gamma[c^*,\mu^*,\bar c^*,\bar \mu^*]
+ \int \scal{ J c^* + K \mu^* } \ee 
defines the 1PI generating functional $\Gamma^{\rm eff}$ associated to
$\Sigma^{\rm eff}$. In this way, we obtain well defined
Slavnov--Taylor identities, which formally take into account the
insertions of the non local operator $G d^\dag \susy A$ and its
transformations. 

After elimination of the shadow fields, for $\alpha=0$, one
has: \be Q \varphi^\A = \susy \varphi^\A - \delta^{\rm gauge}(G
d^\dag \susy A ) \varphi^\A \ee This is the combination of an ordinary
supersymmetry transformation and a gauge transformation with a
field dependent fermionic parameter. This parameter involves the non-local
inverse Faddeev--Popov operator.

This effective $Q$ transformation provides a representation of
supersymmetry on the physical fields, since one has:
\be Q^2 = \L_\kappa + \delta^{\rm gauge}(G d^\dag \L_\kappa A) \ee
For the gauge field, we have: \be Q A = (1 - d_A G
d^\dag) \, \susy A \ee One notices that $1 - d_A G d^\dag$ is a
projector, since $ (1 - d_A G d^\dag)^2 = 1 - d_A G d^\dag $.
Moreover, one has: \be d^\dag (1 - d_A G d^\dag) = 0 \ee Therefore,
after the elimination of the shadow fields, $Q$ can be regarded as
a representation of supersymmetry which is compatible with the
Landau gauge condition \be d^\dag A = 0 \hspace{7mm} \Rightarrow
\hspace{7mm} d^\dag ( A + Q A ) = 0 \ee
Furthermore, one has:
\be
Q\Omega = -Gd^{\dagger }[ \Omega, \susy A-d_A Gd^\dagger \susy A]
\hspace{8mm} Q\bar \Omega =0 \hspace{8mm} Q b = 0
\ee
so one can check that $Q d^\dag d_A \Omega =0$. These remarks indicate
that the Faddeev--Popov determinant admits a supersymmetry in the
absence of the shadow fields, that is:
\be
Q\ {\rm det}\bigl[\, d^\dag d_A \,\bigr]\,\, \delta\,[\,d^\dag A\,] = 0
\ee
However, the non-locality of this symmetry makes it of no practical
use for controlling the quantum theory. In contrast, by keeping the
local dependence on the shadow fields, one can safely use the
conventional tools of local quantum field theory.

%%%%%%%%%%%%%%%%%%%%%%%%%%%%%%%%%%%%%%%%%%%%%%%%%%%%%%%%%%%%%%%%%%
\section{Supersymmetric anomalies}

The standard method to discuss the renormalizability of the theory
to all orders of perturbation theory is as follows.
One assumes that the renormalized generating
functional $\Gamma_{n}$
has been computed at a given order $n$, in such a way
 that it satisfies the Slavnov--Taylor
identities for the $\s$ and $Q$ symmetry and the antighost (and possibly ghost) Ward identities. Then, one
 checks if one can proceed to the next order, with the same conclusion. By
definition, this recursive procedure holds true if there are no
obstructions, i.e. if anomalies are absent. In the present case,
since a manifest invariant regularization framework does not
exist, one must check the absence of anomalies, defined as cohomological non
trivial solutions of the consistency conditions of linearized
operators $\S_{|\Sigma},\, \Q_{|\Sigma},\, \LGg$ and $\LaGg$, for 
ensuring that we can define the supersymmetric theory with all
necessary Ward identities order by order in perturbation theory,
through the introduction of suitable compensating local non
invariant counterterms \cite{PS}. It is thus essential to
 classify the possible solutions of the
consistency conditions of the linearized Slavnov--Taylor operators
that we have constructed, within our class of renormalizable
gauges.

The supersymmetric transformation $\susy$ is nilpotent on the set
$\O $ of gauge-invariant functionals of physical fields
$\varphi^\A$ and supersymmetric parameters. We can consider its
restriction to this set of functionals. This permits us to introduce
a differential complex, $\O^\ast (\susy)$, which is graded by
the shadow number. We shall show that the question of finding the
possible anomalies for the Ward identities $\S,\,
\Q,\, \Gg$ and $\aGg$ is closely linked to that of
finding the cohomology $\H^\ast$ of $\O^\ast (\susy) $.

We will find that the possible anomalies are either elements with
shadow number one in $\H^1$, or pairs built from the $(1,0)$ Adler--Bardeen
anomaly and a $(0,1)$ supersymmetric counterpart. In fact, the latter
possibility only occurs provided we are in a theory where the
following local functional: \be\label{ano}
 {\cal{I}} =
 \int \trace \scal{ F_{\,
\wedge} \susy A_{\, \wedge} \susy A +
 \omega(\varphi) F_{\, \wedge} F} \ee
 is such that $ {\cal{I}} = \susy(\dots)$, that is,
vanishes in $\H^2$.

We must proceed by steps. First, we will
show that the eventual anomalies of the antighost equations can be
removed without spoiling the Slavnov--Taylor identities. Then,
we will be able to restrict to the search of solutions of the ordinary
BRST invariance consistency conditions, the unique solution of which
is the Adler--Bardeen anomaly. Afterward, we will classify the solutions of
the rest of the consistency conditions associated to supersymmetry,
which must be consistent with the BRST invariance. Finally, we will
check that there are no solutions corresponding to the ghost operators
which satisfy all the consistency conditions.

\subsection{ Absence of anomalies associated to the antighost operators}
To demonstrate that there are no anomalies associated with the antighost
operators, it is convenient to assemble the linearized operators
(but the ghost ones) into the following nilpotent differential
operator $\delta$ :
 \be \delta \equiv
\alpha^\a \S_{|\Sigma} + \alpha^\q \Q_{|\Sigma} + \LG[X] +
\LGs[X^\a] + \LGq[X^\q] + \LGqs[X^\qs] + \sigma \ee
 $\alpha^\a$ and $\alpha^\q$ are commuting scalars, $X$ and
$X^\qs$ are anticommuting Lie algebra valued functions and $X^\a$
and $X^\q$, commuting ones. The differential operator $\sigma$ has been introduced in order that: \be
\delta^2 = ({\alpha^\q})^2 \ \P_\kappa \ee so that
$\delta$ is nilpotent, when acting on functionals. The
operator $\sigma$ gives zero on all quantities, but the functions $X^\bullet$, with:
\be\begin{split} \sigma X &= \alpha^\q \L_\kappa X^\q
\\* \sigma X^\qs &= - \alpha^\q X^\a - \alpha^\a X^\q
\end{split}\hspace{10mm} \begin{split}
\sigma X^\a &= - \alpha^\a X - \alpha^\q \L_\kappa X^\qs \\*
\sigma X^\q &= \alpha^\q X
\end{split} \ee
To show that there are no anomalies for the ghost
operators, one can use a standard and general
method \cite{qgs}. It suffices to exhibit that
$\delta$ admits a trivializing homotopy $k$, such that the anticommutator of $k$
and $\delta$ determines the
following counting operator 
\begin{gather}
\{\delta, k\} = e^{ {\textstyle \int} \trace \scal{ - d\bar\Omega \star
 \frac{\delta^L\,\,}{\delta A^\a} + d\bar c \star
 \frac{\delta^L\,\,}{\delta A^\q} + d\bar \mu\star
 \frac{\delta^L\,\,}{\delta A^\qs}}} N e^{- {\textstyle \int} \trace \scal{ - d\bar\Omega \star \frac{\delta^L\,\,}{\delta A^\a} + d\bar c \star
 \frac{\delta^L\,\,}{\delta A^\q} + d\bar \mu\star
 \frac{\delta^L\,\,}{\delta A^\qs}}}\CR
N \equiv \int \trace \Scal{
 b\frac{\delta^L\,\,}{\delta b} + \bar \Omega
 \frac{\delta^L\,\,}{\delta \bar \Omega} + \bar c
 \frac{\delta^L\,\,}{\delta \bar c} + \bar \mu
 \frac{\delta^L\,\,}{\delta \bar \mu} + X \frac{\delta^L\,\,}{\delta
 X} + X^\a \frac{\delta^L\,\,}{\delta X^\a} + X^\q
 \frac{\delta^L\,\,}{\delta X^\q} + X^\qs \frac{\delta^L\,\,}{\delta
 X^\qs}}\label{N}
\end{gather}
This trivializing homotopy is simply :
\be
k \equiv \int \trace \Scal{ b \frac{\delta^L\,\,}{\delta X} + \bar
 \Omega \frac{\delta^L\,\,}{\delta X^\a} + \bar c
 \frac{\delta^L\,\,}{\delta X^\q} + \bar \mu
 \frac{\delta^L\,\,}{\delta X^\qs}}
\ee Therefore, we can always remove the anomalies of
$\S_{|\Sigma}$ and $\Q_{|\Sigma}$ in a way which preserves the
antighost symmetries.
\subsection{Cohomology of the BRST operator $\S_{|\Sigma}$}
We then compute the cohomology of
$\S_{|\Sigma}$ alone. This operator can be split in two parts \be
\S_{|\Sigma} = \Sigma_{\a} + \varsigma \ee where $\Sigma_\a$ is
the operator which acts on any functional
$\F$ of the fields trough the following antibracket \be
\Sigma_{\a} \F = \scal{\Sigma, \F}_\a \equiv \int \Scal{
 \frac{\delta^R \Sigma}{\delta \varphi^\A}
 \frac{\delta^L \F}{\delta \phis_\A} - \frac{\delta^R \Sigma}{\delta \phis_\A}
 \frac{\delta^L \F}{\delta \varphi^\A} + \trace \frac{\delta^R
 \Sigma}{\delta \Omega} \frac{\delta^L \F}{\delta \Omegas} - \trace
 \frac{\delta^R \Sigma}{\delta \Omegas} \frac{\delta^L \F}{\delta \Omega}}
\ee and $\varsigma$ is the $\Sigma$ independent part of $\S_{|\Sigma}$. One
then finds that $\varsigma$ admits a trivializing homotopy $k^\a$
defined by \be k^\a \equiv \int (-1)^\A \phiqs_\A
\frac{\delta^L\,\,}{\delta\phiq} + \int \trace \Scal{ - \Omegaqs
 \frac{\delta^L\,\,}{\delta \Omegaq} - \muq
 \frac{\delta^L\,\,}{\delta \cq} + c \frac{\delta^L\,\,}{\delta \mu}+
 \bar\Omega \frac{\delta^L\,\,}{\delta b} + \bar\mu
 \frac{\delta^L\,\,}{\delta \bar c}}
\ee

The anticommutator of $k^\a$ with $\varsigma$ is the
 counting operator $M$ for all fields but the
physical ones, the ghost $\Omega$ and their sources for the BRST
symmetry $\s$. One can construct perturbatively a functional $V$, such that, the
differential operator $V_\a \equiv \scal{V, \cdot \,}_\a$ satisfies
 \be e^{- V_\a} \S_{|\Sigma} e^{ V_\a} = \Sigma^0_\a +
\varsigma \label{triv}\ee \label{cartan}
Here, $\Sigma^0$ is the
part of $\Sigma$ which is annihilated by $M$ and $k^\a$. This uses the
``Maurer--Cartan" equation :
 \be \varsigma^2
= 0 \hspace{10mm} \varsigma\, \Sigma + \frac{1}{2} \scal{\Sigma,
\Sigma}_\a = 0\ee Eq.(\ref{triv}) implies that the cohomology of
$\S_{|\Sigma}$ is isomorphic to that of $\Sigma^0_\a +
\varsigma$. The latter is itself isomorphic to the cohomology of
$\Sigma^0_\a$ on the set of fields made by the
physical fields, the ghost $\Omega$ and the sources for their $\s$
transformations. This cohomology turns out to be the Adler--Bardeen
anomaly $\mathscr{A}$ in the case of a semi-simple gauge group
\cite{hen}.

\subsection{Supersymmetric anomalies consistent with BRST invariance} 
As announced, we now show that there are two possibilities for the consistent
anomalies of the Ward identities. Either one has a supersymmetric term
associated to the Adler-Bardeen anomaly, or one has a pure
supersymmetric anomaly.
\subsubsection{Possibility for the ``supersymmetric'' Adler--Bardeen anomaly}
Consider the Adler--Bardeen anomaly $\mathscr{A}$. To make it
consistent with the operator $\Q_{|\Sigma}$, we need an integrated
local functional $\mathscr{B}$, possibly depending on all fields
and sources, with ghost number 0, shadow number one and canonical
dimension $\frac{9}{2}$. $\mathscr{B}$ must verify the
consistency conditions \be \S_{|\Sigma} \mathscr{B} + \Q_{|\Sigma}
\mathscr{A} = 0 \hspace{10mm} \Q_{|\Sigma} \mathscr{B} = 0
\label{consistency}\ee

To look for possible solutions, we use the horizontality condition Eq.(\ref{hor1}) and its Bianchi identity, which express
the $\s$ and $Q$ transformation laws. One has :
 \be\label{hor}
 \tilde{F}\equiv (d + \s + Q - i_\kappa)
\scal{ A + \Omega + c} + \scal{ A + \Omega
 + c}^2 = F + \susy A + \omega(\varphi) \ee
The extended
curvature defined by the later equation permits one to use the formal five-dimensional Chern--Simon
identity \be \trace \tilde{F}^3 = \tilde{d} \, \trace \scal{ \tilde{A}
\tilde{F}^2 -
 \frac{1}{2} \tilde{A}^3 \tilde{F} + \frac{1}{10} \tilde{A}^5}
\label{Stora} \ee to determine both the Adler--Bardeen
anomaly and its supersymmetric counter part. The term with
 ghost and shadow numbers $(2,0)$ shows that the Adler--Bardeen
 anomaly satisfies the consistency condition for $\s$ :
\be \mathscr{A} \equiv
\int \trace d\Omega \scal{ A d A + \frac{1}{2} A^3} \hspace{10mm}
\s \mathscr{A} = 0 \ee
The term with ghost and shadow
numbers $(1,1)$ gives a solution for the consistency condition \be
\s \mathscr{B}^\zero + Q \mathscr{A} = 0 \label{cons}\ee
It is defined by
\be \mathscr{B}^\zero \equiv \trace \Bigl( d c \scal{ A d A +
 \frac{1}{2} A^3} + \scal{ [A, F] - \frac{1}{2} A^3} \susy A\Bigr) \ee
The term with ghost and shadow numbers $(0,2)$ gives
the breaking of the consistency condition
 $Q \mathscr{B}=0$ for this solution. Indeed, one gets :
\be Q \mathscr{B}^\zero = 3 \int \trace \scal{ F_{\, \wedge} \susy
A_{\,
 \wedge} \susy A + \omega(\varphi) F_{\, \wedge} F} \ee
To obtain a consistent $\mathscr{B}$, we must add a
$\S_{|\Sigma}$-closed term $\mathscr{B}^\un$ to $\mathscr{B}^\zero$, 
 in order to have a $\Q_{|\Sigma}$-closed functional $\mathscr{B}$, which 
preserves Eq.~(\ref{cons}). Since $Q 
\mathscr{B}^\zero$ is not $\S_{|\Sigma}$-exact, $\mathscr{B}^\un$ must
be in the cohomology of $\S_{|\Sigma}$. Therefore $\mathscr{B}^\un$
is a gauge-invariant functional on the physical fields.
Because the transformation $\Q_{|\Sigma}$ acts as $\susy$ on
 gauge-invariant functionals, we obtain the following result.
 A functional $\mathscr{B}$, which satisfies the consistency equations (\ref{consistency}), where
 $\mathscr{A}$ is the Adler--Bardeen anomaly, exists, if and only if,
 there is a gauge-invariant local functional
$\mathscr{B}^\un$, for which the following
equation holds
\be \susy \mathscr{B}^\un = - 3 \int \trace \scal{
F_{\, \wedge} \susy A_{\, \wedge} \susy A + \omega(\varphi) F_{\,
\wedge} F} \ee
For the case $\mathcal{N}=2,\, 4$, the right hand side in the last equation can be identified as a cohomology class of the operator
$\susy$. It can be associated to Donaldson--Witten invariants \cite{in
  preparation}. This gives an algebraic proof of the absence of Adler--Bardeen
anomaly in extended supersymmetry. This feature is usually
attributed to the fact that there are no chiral fermions in such
theories.

For the case $\mathcal{N}=1$, one can construct the following
antecedent:\footnote{For $\mathcal{N}=1$ in 
Minkowski space time \be \susy A_\mu = -i
\scal{\overline{\epsilon} \gamma_\mu \lambda} \hspace{10mm} \susy
\lambda = \scal{ \baa F + \gamma_5 H } \epsilon \hspace{10mm} \susy
H = i \scal{\overline{\epsilon} \gamma_5 \baaa D
 \lambda } \ee
with \be (\susy)^2 = i \scal{\overline{\epsilon} \gamma^\mu
\epsilon} \scal{
 \partial_\mu + \delta^{\rm gauge} (A_\mu)}\ee }
\be \susy \frac{1}{4} \trace \scal{\overline{\epsilon} \gamma^\mu \lambda}
\scal{\overline{\lambda} \gamma_5 \gamma_\mu \lambda} = \frac{1}{2}
\varepsilon^{\mu\nu\sigma\rho} \trace
F_{\mu\nu}\scal{\overline{\epsilon} \gamma_\sigma
 \lambda}\scal{\overline{\epsilon} \gamma_\rho \lambda} \ee
%This confirms the fact that in $\mathcal{N}=1$ super-Yang--Mills
%the matter representation of the gauge group must be chosen
%carefully in order to avoid the Adler--Bardeen anomaly.
Thus, gauge anomalies can exist in this case. They can be detected as gauge anomalies and as anomalies in the supersymmetry Noether current.

\subsubsection{Purely supersymmetric anomalies}
Since the Adler--Bardeen is the only possible anomaly for the
$\S_{|\Sigma}$ operator, other potential anomalies for
$\Q_{|\Sigma}$ must be in the kernel of $\S_{|\Sigma}$.
In other word, we may have a term ${\mathscr{C}}^\gra{0}{1}$,
with
\be
\S_{|\Sigma}\mathscr{C}=0 \hspace{10mm} \Q_{|\Sigma}\mathscr {C}=0
\ee
As a
result of \cite{hen}, any element of the kernel of $\S_{|\Sigma}$
of ghost number zero can be decomposed into a gauge-invariant
functional in the physical fields and a $\S_{|\Sigma}$-exact term.
Thus, one computes the cohomology of $\Q_{|\Sigma}$ in the
image of $\S_{|\Sigma}$. Any such element must be of the following
form
\begin{multline}
\mathscr{C}^\im = \S_{|\Sigma} \int \biggl(\bigl<\bar \Omega ,\,
\lf{0}{1}{\frac{5}{2}}\bigr> +
 \bigl< \bar \mu, \lf{0}{2}{3}\bigr> + \bigl< \phis_\A,
 \ff{0}{1}{[a]+\frac{1}{2}}^\A\bigr> +
\bigl< \phiqs_\A, \gf{0}{2}{[a]+1}^\A \bigr> \\*+ \bigl< \Omegas,
\ff{1}{1}{\frac{1}{2}} \bigr>+ \bigl< \Omegaqs, \gf{1}{2}{1}\bigr> +
\bigl< \Omegaq,
\hf{0}{2}{1}\bigr> + \bigl< \muq, \hf{0}{3}{\frac{3}{2}}\bigr> \biggr)
\end{multline}
Here $\lf{g}{s}{d} $, $\ff{g}{s}{ d} $, $\gf{g}{s}{ d} $ and
$\hf{g}{s}{d} $ are local functions of the fields, where the sub-index $d$ is the canonical dimension. The explicit form
of $\mathscr{C}^\im$ is
\begin{multline}
\mathscr{C}^\im = \int \biggl(\bigl<b , \lf{0}{1}{\frac{5}{2}}\bigr>
-\bigl<\bar \Omega , \S_{|\Sigma} \, \lf{0}{1}{\frac{5}{2}}\bigr> +
 \bigl< \bar c, \lf{0}{2}{3}\bigr> + \bigl< \bar \mu , \S_{|\Sigma}\,
 \lf{0}{2}{3}\bigr> \\*+ \bigl< \frac{\delta^R \Sigma}{\delta \varphi^\A} ,
 \ff{0}{1}{[a]+\frac{1}{2}}^\A\bigr> + (-1)^\A \bigl< \phiq_\A,
 \gf{0}{2}{[a]+1}^\A \bigr> -(-1)^\A \bigl< \phis_\A, \S_{|\Sigma}
 \, \ff{0}{1}{[a]+\frac{1}{2}}^\A\bigr> + (-1)^\A \bigl< \phiqs_\A,
 \S_{|\Sigma} \, \gf{0}{2}{[a]+1}^\A \bigr> \\*+ \bigl<
 \frac{\delta^R \Sigma}{\delta \Omega}
 , \ff{1}{1}{\frac{1}{2}} \bigr>- \bigl< \Omegaq, \gf{1}{2}{1} -
 \S_{|\Sigma}\,\hf{0}{2}{1} \bigr>+ \bigl< \Omegas,
\S_{|\Sigma} \, \ff{1}{1}{\frac{1}{2}} \bigr> - \bigl< \Omegaqs, \S_{|\Sigma}
\, \gf{1}{2}{1}\bigr>\\* - \bigl< \cq, \hf{0}{3}{\frac{3}{2}}\bigr>
-\bigl< \muq,
\S_{|\Sigma} \, \hf{0}{3}{\frac{3}{2}}\bigr> \biggr)
\end{multline}
The consistency conditions imply that : \be
\G_\bullet \mathscr{C}^\im = 0 \ee Thus :\be \lf{0}{1}{\frac{5}{2}}
= \lf{0}{2}{3} = 0 \ee On the other hand, the condition
$\Q_{|\Sigma} \mathscr{C}^\im = 0$ yields at zero-th order in the sources :
\begin{multline}
 \int\biggl( \bigl< \frac{\delta^L \Sigma}{\delta \varphi^\A} ,
 \Q_{|\Sigma} \, \ff{0}{1}{[a]+\frac{1}{2}}^\A + \gf{0}{2}{[a]+1}^\A
 \bigr> + \bigl< \frac{\delta^L \Sigma}{\delta \Omega}, \Q_{|\Sigma}
 \, \ff{1}{1}{\frac{1}{2}} + \gf{1}{2}{1}- \S_{|\Sigma}\, \hf{0}{2}{1}
 \bigr> \\*+ \bigr< \frac{\delta^L \Sigma}{\delta c},
 \hf{0}{3}{\frac{3}{2}}\bigr> - \bigl< \frac{\delta^L \Sigma}{\delta
 \mu}, \S_{|\Sigma}\,\hf{0}{3}{\frac{3}{2}}\bigr> \biggr) =
\mathcal{O}(\varphi^\bullet) \label{sym}
\end{multline}

 This equation is solved, either when each term vanishes, or when
the action $\Sigma$ admits an accidental symmetry, with ghost and
shadow numbers $(0,2)$, and canonical dimension one, which only acts on
the physical fields, $\Omega$, $c$ and $\mu$. It is reasonable to
assume that this later possibility never arises. Therefore, one must have :
 \bea \label{decom}
 \Q_{|\Sigma}\, \ff{0}{1}{[a]+\frac{1}{2}}^\A + \gf{0}{2}{[a]+1}^\A &=&
 \cf{0}{2}{[a]+[b]-3}^{\A\B} \frac{\delta^R \Sigma}{\delta \varphi^\B}
 + \mathcal{O}(\varphi^\bullet) \CR
\Q_{|\Sigma}\, \ff{1}{1}{\frac{1}{2}} + \gf{1}{2}{1}- \S_{|\Sigma}
\hf{0}{2}{1} &=& \mathcal{O}(\varphi^\bullet) \CR
\hf{0}{3}{\frac{3}{2}} &=& \mathcal{O}(\varphi^\bullet)
\eea
This property tells us that $\mathscr{C}^\im$ is trivial at
zeroth order in the sources $\varphi^\bullet$, that is :
\be \mathscr{C}^\im = \S_{|\Sigma} \Q_{|\Sigma} \int \Scal{ - (-1)^\A \bigl<
 \phiqs_\A , \ff{0}{1}{[a]+\frac{1}{2}}^\A \bigr> + \bigl< \Omegaqs,
\ff{1}{1}{\frac{1}{2}} \bigr> } + \mathcal{O}(\varphi^\bullet) \ee
Due to the power counting, only the functions
$\ff{0}{1}{[a]+\frac{1}{2}}^\A$ and $\gf{0}{2}{[a]+1}^\A$ can
 depend on the external sources. Moreover, this dependence must be
 linear. One can then compute that $\mathscr{C}^\im$ must be trivial at all order in the sources, extending the
equations (\ref{sym},\ref{decom}). We thus obtain the desired result that the cohomology of
$\Q_{|\Sigma}$ in the image of $\S_{|\Sigma}$ is empty.

We have thus shown that, the possible purely supersymmetric anomalies
$\mathscr{C} $ are the elements of $\H^1$ with canonical dimension
four. In fact, $\H^1$ depends on the given supersymmetric theory. As a
direct consequence of results in \cite{maggiore}, $\H^1 = \{0 \}$ in
$\mathcal{N}=2$ super Yang--Mills. Preliminary computations show that
there are no supersymmetric anomalies in $\mathcal{N}=4$ superYang--Mills, in
agreement with \cite{conform}. 

To summarize, anomalies
for the linearized operators $\LGg$, $\S_{|\Sigma}$ and
$\Q_{|\Sigma} $ must be, either an element of $\H^1$, or the
Adler--Bardeen anomaly $\mathscr{A}$, provided that a
corresponding counterpart $\mathscr{B}$ exists.

\subsection{Solutions of the consistency conditions associated to the
 ghost operators}

In the Landau gauge, there are additional symmetries, expressed by the
ghost operators $ \aGg$, whose homogeneous parts are the differential
operators $\LaGg$. This gauge has many advantages, but
 one must check that there is no anomaly for the operators $\LaGg$.
 From the previous computation, one can check that the possible anomalies
 for $\LGg$, $\S_{|\Sigma}$ and $\Q_{|\Sigma}$ are left invariant by
 the four operators $\LaGg$'s. Therefore, the computation of possible anomalies
 of the ghost operators reduces to a relatively simple system of
 consistency equations. Moreover, because the operators $\LaGg$
 decrease the ghost and shadow numbers, their anomalies
 $\Delta^\bullet$ must depend on the anti-ghosts, the anti-shadows, and
 the sources, which have large canonical dimensions. By inspection, one
 finds that the consistency conditions of $ \Delta^\bullet$ have no
 local solutions with the required canonical dimension.

 Therefore, one can use the ghost
Ward identities of the Landau gauge, and the anomaly problem is
the same in this gauge as for all other values of $\alpha$. The
advantage of the Landau gauge, is that it simplifies the proof of
the stability of the action. Most importantly, it is a preferred
gauge for demonstrating non-renormalization theorems~\cite{PS,gSor}.

%\begin{gather} \bigstar\CR
%\bigstar \hspace{7mm} \bigstar \nonumber\end{gather}

%
% In summary, the cohomological problem of the
%characterization of the anomalies related to the whole set of Ward
%identities reduces to a cohomological problem which makes use only
%of the classical fields and of the supersymmetric transformations.
%The eventual anomalies which could arise are: gauge invariant
%functionals of ghost and shadow numbers
%$(0,1)$ belonging to the cohomology of $\susy$ and a
%supersymmetric extension of the Adler--Bardeen anomaly which is
%consistent if and only if the $\susy$-closed functional
%\be \int \trace \scal{ F_{\, \wedge} \susy A_{\,
% \wedge} \susy A + \omega(\varphi) F_{\, \wedge} F} \ee
%is $\susy$-exact.
\section{General local solution of the Slavnov--Taylor identities}
In this section we assume that the anomalies mentioned in the last
section do not exist in the supersymmetric theory we are dealing
with, or at least that these anomalies do not arise in
perturbation theory. For $\mathcal{N}=2,4$ no anomaly can arise, but
for $\mathcal{N}=1$, the consistency conditions may have solutions, and the coupled hypermultiplet
must be carefully chosen.

To prove the stability of the
action, i.e. to show that the ambiguities related to the local
invariant counterterms, which can be freely added at each order of
perturbation theory, do not exceed the number of parameters of the
classical starting action, we must compute the most general
integrated local polynomial, with ghost and shadow numbers $(0,0)$
and canonical dimension four, which satisfies all Ward identities. We
will work in the Landau gauge, $\alpha=0$.

Because of the ghost equations and power counting, the source
dependence of the action must be linear, but for possible quadratic
terms in the $\phiq_\A$. However, since the Slavnov--Taylor identities mix the
different sources ($\phiq_\A$, $\phis_\A$ and $\phiqs_\A$), such quadratic
terms must be absent. Therefore, the most general solution
$\Sigma^\R$ of the Slavnov--Taylor identities must have a linear
dependence on all sources. It can be written as follows :
\begin{multline}
\Sigma^\R = S^\R[\varphi^\A, \Omega, \bar\Omega, c, \bar c, b, \bar \mu,
\mu] + \int (-1)^\A \Scal{ \phis_\A \Delta^\A_\a + \phiq \Delta^\A_\q
 + \phiqs \Delta^\A_\qs} \\*
+ \int \trace \Scal{ -\Omegas \Delta_\a - \Omegaq \Delta_\q - \Omegaqs
 \Delta_\qs + \muq \Lambda_\qs - \cq \Lambda_\q}
\end{multline}
 Here, the $\Delta_\bullet$'s are local polynomials in the quantum fields, which can be used to define differential operators $\s^\R$ and $Q^\R$.
 \begin{gather} \Delta^\A_\a = \s^\R \varphi^\A
\hspace{10mm} \Delta^\A_\q = Q^\R \varphi^\A\CR \Delta^\A_\qs =
\s^\R Q^\R \varphi^A \CR \Delta_\a = \s^\R \Omega \hspace{10mm}
\Delta_\q = Q^\R \Omega\CR \Delta_\qs = \s^\R Q^\R \Omega\CR
\Lambda_\qs = Q^\R \mu \hspace{10mm} \Lambda_\q = Q^\R c
\end{gather}
 The Slavnov--Taylor identities imply that $ S^\R$ must be invariant
 under the action of $\s^\R$ and $Q^\R$, and that these latter must
 verify
\begin{gather}
{\s^\R}^2 = 0 \hspace{10mm} {Q^\R}^2 = \L_\kappa \CR
\{\s^\R, Q^\R\} = 0
\end{gather}
$\s^\R$ and $Q^\R$ acts as $\s$ and $Q$ when these ones act
linearly. Modulo a rescaling of ghosts and shadows $\Omega$, $\mu$ and $c$
as well as of the structure constants of the gauge group, it turns
out that the operators $\s^\R$ and $Q^\R$ must have the same form
of $\s$ and $Q$. Thus, the gauge symmetry can only be
renormalized multiplicatively, and supersymmetric transformations
must be renormalized in the most general possible way, which is
compatible with power counting and their closure on a renormalized
gauge transformation. \be\label{dist} {\susy_\R}^2 =
\delta^{\mathrm{gauge}}(\omega^\R(\varphi) + i_\kappa A) +
\L_\kappa \ee Then, the renormalized action decomposes into a part
belonging to the cohomology of $\s^\R$ and into a $\s^\R$-exact
part. Because $\s^\R$ and
$\s$ are identical modulo rescalings of ghosts and the coupling constant,
 the former term is a gauge
invariant local functional which depends only on the fields
$\varphi^\A$. It must also be invariant under the
renormalized supersymmetry. Since the transformations are the same
on the fields of negative ghost number, the $\s^\R$-exact part
must be $Q^\R$-exact in order to be $Q^\R$ invariant. The antighost
equations then fix this part to be the one of the classical action
\be \s^\R Q^\R \int \trace \bar \mu d \star A \ee To go further in
determining $\omega^\R(\varphi) $, one must look at the details of
the particular supersymmetric theory. Therefore, the conclusion
is that in a supersymmetric theory with a supersymmetric algebra
which closes off-shell, and which has no anomaly, the action is
stable under radiative corrections if and only if, the most
general supersymmetric algebra defined on the set of fields from
Eq.~({\ref{dist}}), can be related to the standard one by linear
redefinitions of the fields. In fact, this is the case in all super
Yang--Mills theories.

\section{Interpolating gauges}
 
\label{nonSusy} In this section, we discuss the extension of
the Slavnov--Taylor identity associated to supersymmetry to an
arbitrary value of the gauge parameter $\zeta$. As already mentioned, this is 
achieved by introducing an anticommuting parameter $\z$ which
allows one to vary the gauge parameter $\zeta$, while maintaining
the desired Slavnov--Taylor identity.

The anticommuting parameter $\z$ is such that
$Q\z=\zeta$, $Q \zeta=0$, and $\s \zeta=0$, $\s \z=0$. For $\zeta\neq
0$, one has a non-vanishing $Q$-variation of the gauge-fixing term.
Using the gauge-fixing term eq.(\ref{Zsource}), one has, however, an
extended Slavnov--Taylor identity: 
\be
\hspace{10mm} \Q^{'}( \Sigma ') \equiv \Q( \Sigma ') + \zeta
\frac{\partial \Sigma}{\partial \z} = 0 \ee 
The antighost Ward identities can also be extended to a
generic value of the gauge parameters $\zeta$ and $\z$. In the present case,
these Ward identities generalize to
\begin{gather}
\G(\F) \equiv \int \trace X \Scal{ \frac{\delta^L \F}{\delta b} -
 d\star A + \alpha \star b- \z d \star\frac{\delta^L \F}{\delta
 A^\q}+\z d \star d c } \CR
\Gs (\F) \equiv \int \trace X \Scal{\frac{\delta^L
 \F}{\delta \bar \Omega} + d \star \frac{\delta^L \F}{\delta
 \As}+ \z d \star \frac{\delta^L \F}{\delta A^\qs}+ \z d \star d \mu}
\hspace{70mm} \CR
 \Gq (\F) \equiv \int \trace X \Scal{\frac{\delta^L
 \F}{\delta \bar c} - (1-\zeta) d \star \frac{\delta^L \F}{\delta
 \Aq}-\zeta d\star d c - \alpha \star \L_\kappa \bar c+ \z d \star d
\frac{\delta^L \F}{\delta c^\q}
 + \z \L_\kappa d \star A} \CR
\Gqs (\F) \equiv \int \trace X \Scal{\frac{\delta^L
 \F}{\delta \bar \mu} + (1-\zeta) d \star \frac{\delta^L \F}{\delta
 \Aqs} - \zeta d\star d \mu-\z d \star d \frac{\delta^L \F}{\delta
 \mu^\q }- \z \L_\kappa d \star \frac{\delta^L \F}{\delta A^\a}}
\end{gather}
The consistency conditions for the operators $\S,\,\Q$ and
$\Gg$ remain the same. For the classification of anomalies,
 one can repeat the same arguments as previously, with $\zeta \neq
 0$. The only change is that, the
translation operator :
\be e^{ {\textstyle \int} \trace \scal{ -
d\bar\Omega \star
 \frac{\delta^L\,\,}{\delta A^\a} + d\bar c \star
 \frac{\delta^L\,\,}{\delta A^\q} + d\bar \mu\star
 \frac{\delta^L\,\,}{\delta A^\qs}}}
\ee
which appears in (\ref{N}), must be modified to
\be
e^{ {\textstyle \int} \trace \scal{ - d\bar\Omega \star
 \frac{\delta^L\,\,}{\delta A^\a} + ((1-\zeta) d\bar c + \z db) \star
 \frac{\delta^L\,\,}{\delta A^\q} + ( (1-\zeta) d\bar \mu - \z d \bar
 \Omega)\star \frac{\delta^L\,\,}{\delta A^\qs} - \z d \star d \bar c
 \frac{\delta^L \,\,}{\delta \cq} - z d \star d \bar \mu\frac{\delta^L
 \,\,}{\delta \muq} }}
\ee
 When checking the stability of the action, one still finds that the 
 gauge-fixing action $\s \Uppsi$, which depends on the parameters $\zeta$ and
 $\z$, is stable. However, we have not been able to derive ghosts Ward
identities and prove that the action depends linearly in the sources,
even in the Landau gauge. As a matter of fact, more computations are
needed, to obtain the stability of the whole action, including the sources.

However, ghost Ward identities are in fact not necessary to remove
ambiguities. What matters is that the supersymmetric gauge,
$\zeta=0$, can be analytically linked to the other gauges $\zeta\neq
0$, and, in particular, to the ordinary Faddeev--Popov 
gauge-fixing. As we have just seen, one can extend all the required
Ward identities necessary to define the theory in the
interpolating gauge. The Green functions of quantities which
are left invariant by the operators $\S_{|\Sigma}$ are independent from
the parameter $\zeta$ and $\z$. The supersymmetric gauge $\zeta=0,\
\z=0$ gives the simplest gauge for explicit computations and check the
Slavnov--Taylor identities at each order in perturbation theory.  For
$\zeta\neq 0$, one must keep the $\zeta,\ \z$ dependence, in order to
control supersymmetric covariance.

\section{ An example: the case of $\mathcal{N}=2$ super-Yang--Mills}
\label{exemple}
 In order to illustrate our general formalism, let us
consider the example of $\mathcal{N}=2$ super-Yang--Mills in
Euclidean space. In this case the supersymmetry algebra closes
off-shell with the introduction of an auxiliary field $H^i$ in the
adjoint representation of the $SU(2)$ R-symmetry group. The theory
contains a gauge field $A$, a $SU(2)$-Majorana fermion $\lambda$
(we do not write the R-symmetry index), one scalar $\phi$ and one
pseudo-scalar $\phi^5$. The classical action is 
\begin{multline}
\int d^4 x \trace \biggl ( -\frac{1}{4} F_{\mu\nu} F^{\mu\nu} +
\frac{1}{2} H_i H^i + \frac{1}{2} D_\mu \phi^5 \,
D^\mu \phi^5 - \frac{1}{2} D_\mu \phi \, D^\mu \phi -
\frac{i}{2} \scal{\overline{\lambda} \baaa D \lambda} \biggr .\\*\biggl .
\frac{1}{2} [\phi^5, \phi]^2 + \frac{1}{2} \scal{\overline{\lambda}
 \gamma_5 [\phi^5, \lambda]} - \frac{1}{2}\scal{\overline{\lambda} [\phi,
 \lambda]} \biggr )
\end{multline}
and the supersymmetric algebra is :
\bea
\susy A_\mu &=& -i\scal{\overline{\epsilon}\gamma_\mu\lambda}\CR
\susy\phi^5 &=&
-\scal{\overline{\epsilon}\gamma_5\lambda}\hspace{10mm} \susy \phi =
-\scal{\overline{\epsilon}\lambda}\CR
\susy \lambda &=& \baa F\epsilon + \gamma_5[\phi^5,\phi]\epsilon  
-i\scal{\gamma_5\baaa D\phi^5+\baaa D\phi}\epsilon
+H_i\tau^i\epsilon\CR
\susy H^i &=& \Bigl(\overline{\epsilon}\tau^i\scal{-i\baaa
 D\lambda +\gamma_5[\phi^5,\lambda]-[\phi,\lambda]}\Bigr)\CR
\eea 
 It closes as follows \be (\susy)^2 =
\delta^{\mathrm{gauge}}\scal{(\overline{\epsilon} [ 
 \gamma_5 \phi^5 - \phi ] \epsilon ) -i (\overline{\epsilon}
 \gamma^\mu \epsilon) A_\mu} - i (\overline{\epsilon} \gamma^\mu
 \epsilon) \partial_\mu \ee
 
The supersymmetric gauge fixing-term is : 
\begin{multline}
- \s Q \int d^4 x \trace \bar \mu \scal{\partial^\mu A_\mu + \frac{\alpha}{2} b}
=\\* \int d^4 x \trace \biggl(-b \partial^\mu A_\mu - \frac{\alpha}{2} b^2 + \bar
\Omega \partial^\mu D_\mu \Omega - \bar c \partial^\mu D_\mu c - \bar
\mu \partial^\mu D_\mu \mu \\*- i \frac{\alpha}{2} (\overline{\epsilon}
\gamma^\mu \epsilon) \bar c \partial_\mu \bar c + i \bar c
(\overline{\epsilon} \, \ba \partial \lambda) + \partial_\mu \bar \mu
\scal{ [D^\mu \Omega, c] +
 i(\overline{\epsilon} \gamma^\mu [\Omega, \lambda])} \biggr)
\end{multline}
 We notice that this gauge-fixing term gives
 propagators between the shadow and the fermion
fields, which depends on the supersymmetric parameters, as follows : \be
 \left<c \ \lambda \right>_0 = - \frac{\epsilon}{p^2}
 \hspace{10mm} \left<c \ c \right>_0 =
 (1-\alpha)\frac{(\overline{\epsilon} \ba
 p \epsilon )}{(p^2)^2}
\ee
 We checked that, no infra-red problem , which cannot be cured, is
implied by the apparently infra-red dangerous propagator $ \left<c
 \ c \right>_0$. The dependence on the supersymmetric parameters
$\epsilon$ of the propagators has no physical consequences for the
observables, since the latter can be considered as
gauge parameters. Their appearance in the propagators can be
handled
 as that of the parameter $\alpha$
 in the
gauge propagator:
 \be \left< A_\mu \
A_\nu \right>_0 = - \frac{ \delta_{\mu\nu} -
 (1-\alpha) \frac{p_\mu p_\nu}{p^2}}{p^2} \ee

\section{Conclusion}
An important component of this paper is the introduction of a new sector in super-Yang--Mills theories, namely that of shadow fields. It defines new
 gauge-fixing terms that
depend on a more general family of gauge parameters. Some of them can be
identified as the ``constant ghosts'' for supersymmetry
transformations. The preserved BRST symmetry allows one to define
supersymmetric covariant and gauge-invariant observables.
Moreover, the Slavnov--Taylor identities imply that the
correlation functions of the observables do not depend on these
parameters. The ordinary Faddeev--Popov gauge-fixing, which is not
supersymmetric, is obtained for a given choice of these
parameters. Other choices give an explicitly supersymmetric
gauge-fixing. This provides a proof that supersymmetry and gauge
invariance of physical observables can be safely maintained, even
in the simplest non-supersymmetric gauges.

%%%%%%%%%%%%%%%%%%%%%%%%%%%%%%%%%%%%%%%%%%%%%%%%%%%%%%%%%%%%%%%%%%%%%%%%%%%%%

\section*{Acknowledgments.}
 We thank Raymond Stora for numerous discussions on the subject. 
 
 This work was partially supported under the contract ANR(CNRS-USAR) \\ \texttt{no.05-BLAN-0079-01}.

S.P. Sorella thanks the Conselho Nacional de Desenvolvimento
Cient\'{\i}fico e Tecnol\'{o}gico (CNPq-Brazil), the FAPERJ and
the SR2-UERJ for financial support.

\end{document}